\documentclass[a4paper,11pt]{article}

\usepackage{soul}
\usepackage{color}
\usepackage[singlelinecheck=off,center]{caption}
\usepackage[T2A]{fontenc}
\usepackage[utf8x]{inputenc}
\usepackage[warn]{mathtext}
\usepackage{indentfirst}
\usepackage[fleqn]{amsmath}
\usepackage{amsthm,amsfonts,amssymb,amscd}
\usepackage{hyperref}
\usepackage{bm}
\usepackage{icomma}
\usepackage{mathrsfs}
\usepackage{cite}

\usepackage{physics}

\renewcommand{\vec}{\mathbf}

\begin{document}

\title{Quantum process in probability representation of quantum mechanics}
\author{Y. V. Przhiyalkovskiy \\ \\ \textit{\small Kotelnikov Institute of Radioengineering and Electronics (Fryazino Branch)}\\ \textit{\small of Russian Academy of Sciences,}\\ \textit{\small Vvedenskogo sq., 1, Fryazino, Moscow reg., 141190, Russia} \\ \small e-mail: yankus.p@gmail.com}
\date{}

\maketitle

\begin{abstract}
In this work, the operator-sum representation of a quantum process is extended to the probability representation of quantum mechanics. It is shown that each process admitting the operator-sum representation is assigned a kernel, convolving of which with the initial tomogram set characterizing the system state gives the tomographic state of the transformed system. This kernel, in turn, is broken into the kernels of partial operations, each of them incorporating the symbol of the evolution operator related to the joint evolution of the system and an ancillary environment. Such a kernel decomposition for the projection to a certain basis state and a Gaussian-type projection is demonstrated as well as qubit flipping and amplitude damping processes.
\end{abstract}

%\keywords{symplectic tomography, probability representation of quantum mechanics, quantum process, operator-sum representation}

\maketitle

\section{Introduction}

The probability representation of quantum mechanics \cite{Mancini1997}, otherwise called symplectic tomography, offers a new formulation of quantum mechanics and allows us to look at this theory in a new way. The background of this formalism is formed by the representation of a quantum state by a two-parameter set of probability densities, or tomograms, which are indeed measurable. Actually, the Hilbert space, which is the primary structure of any quantum system in the traditional formulation, is replaced by the space of probability distributions. Operators acting on vectors in Hilbert space are unambiguously mapped to $C$-functions in a special way. In particular, a set of tomograms specifying the state is nothing but the result of such a mapping applied to a density matrix. To deal with functions associated with operators, so-called operator symbols, without losing the inherent to quantum mechanics noncommutativity, they are furnished with a star product formalism. All this machinery eventually brings the conventional quantum formalism into the domain of familiar functions and is well covered in the literature \cite{Ibort2009, Manko2021, Fedorov2013}.

Although the probability representation of quantum mechanics is studied mainly theoretically, it is nevertheless promising for applied physics. To obtain an optical tomogram in quantum optics, the well-developed method of balanced homodyne detection is widely used \cite{Lvovsky2009}. A symplectic tomogram is a natural generalization of an optical tomogram and, in comparison with it, contains redundant information. In return, however, recovering the quantum state in the form of a Wigner function from a set of symplectic tomograms is simpler than from optical tomograms \cite{Facchi2010}. It should be emphasized that symplectic tomograms, unlike the wave function or Wigner function, are acquired directly from the experiment. Hence, staying completely within the tomographic formalism, it becomes no need for any further calculations to reconstruct the state. Methods of the symplectic tomography can be focused not only directly on reconstructing purposes, in particular, using the framework of symmetric informationally complete positive operator-valued measure (SIC-POVM) \cite{Filippov2010}, but also can be applied in quantum state engineering and control, in designing components for quantum computing or quantum information applications \cite{Fedorov2013-2}, in quantum process simulation \cite{Arkhipov2003-2, Lozovik2004}.

When we consider real systems, we generally must take into account their coupling with the environment. The most general concept underlying the mathematical framework of describing open systems is a quantum process (quantum channel) which essentially comprises a completely positive trace-preserving map of the density matrix. Although in general not all, most of these maps admit the operator-sum representation \cite{Kraus1971}. Within this framework, the process is decomposed to a set of partial operations that occur with corresponding probabilities, each operation being assigned with a Kraus operator. Further, taking the Markovian approximation leads to the prominent Gorini-Kossakowski-Sudarshan-Lindblad (GKSL) equation, which describes the memoryless evolution of the density matrix \cite{Gorini1976, Lindblad1976}. The natural question is eventually arises: how the above construction enters the tomographic picture?

The application of the methods of symplectic tomography to complement the theory of open quantum systems is present in literature but, though, insufficiently. The most straightforward way to track the tomographic dynamics of an open system is to use traditional formalism and then transform the results into tomograms \cite{Rohith2015, Thapliyal2016}. In Ref. \cite{Bellomo2009}, the authors use the symplectic tomography technique just to reconstruct the parameters of the specific GKSL-type equation, in Ref. \cite{Bellomo2010} this procedure was extended to the convolutionless non-Markovian case. However to stay within the probability representation of quantum mechanics, the differential equation which governs the tomogram evolution is extracted straightforwardly from the GKSL equation by transforming operators to their symbols \cite{Mancini1997, Manko2021}. This equation resembles the Fokker-Plank one, which originally describes the probability density dynamics in statistical physics. Nevertheless, it is essential to emphasize that the evolution equation derived in the tomographic picture inherits the Markovian character of the base GKSL equation. Concerning non-Markovian dynamics, however, no general approaches such as, for example, the Nakajima-Zwanzig formalism \cite{Nakajima1958, Zwanzig1960}, have been elaborated within the symplectic tomography yet.

The aim of the current work is to elaborate the counterpart of the operator-sum representation of quantum processes in terms of the probability representation of quantum mechanics. This issue so far been addressed only partially \cite{Manko2006}, however a general formalism to describe quantum processes is lacking. We show, that a process kernel resembling a tomogram propagator can be associated with a quantum process. For a non-unitary quantum process, this kernel is broken into several partial kernels in the same fashion as it occurs in the conventional operator-sum representation. Note, that this is the process decomposition that was encountered in \cite{Przhiyalkovskiy2021} for a system undergone a non-selective continuous measurement. The constraints imposed to operator symbols and partial operation kernels to be appropriate for describing a general quantum process are derived. To illustrate the operator-sum representation from the viewpoint of the probability representation of quantum mechanics, the von Neumann measurement model and projective measurement are examined in detail and supplemented by examples of flipping and amplitude damping of a qubit.

\section{Symplectic tomography}

To begin with, let us take a brief survey of the main aspects of symplectic tomography. Define a self-adjoint operator $\hat{X} = \mu \hat{q} + \nu \hat{p}$ being the general linear combination of the coordinate and momentum operators. Actually, this map is a "coordinate" part of a general symplectic transformation of the coordinate and momentum operators, namely, the linear transformation that preserves the coordinate-momentum commutator. Following the original idea of symplectic tomography \cite{Mancini1997}, a tomogram is a probability distribution of the coordinate $\hat{X}$, parameterized by two real variables $\mu$ and $\nu$. 

For a further review of the symplectic tomography, we will employ the most convenient formalism involving the dequantizer and quantizer operators \cite{Manko2021}. Let $\vec{x} = (X, \mu, \nu)$ be a vector containing three real variables. Define the operators $\hat{\mathcal{U}}(\vec{x})$ and $\hat{\mathcal{D}}(\vec{x})$, referred to as the dequantizer and quantizer, respectively, as
\begin{equation}
	\begin{aligned}
		& \hat{\mathcal{U}}(\vec{x}) = \delta(X - \mu \hat{q} - \nu \hat{p}), \\
		& \hat{\mathcal{D}}(\vec{x}) = \frac{1}{2\pi} \exp \left( iX - i \mu \hat{q} - i\nu \hat{p} \right), \\
	\end{aligned}
	\label{eq:UD}
\end{equation}
which satisfy the consistency condition $\operatorname{Tr} \{ \hat{\mathcal{U}}(\vec{x}_1) \hat{\mathcal{D}}(\vec{x}_2) \} = \delta(\vec{x}_1 - \vec{x}_2)$. Here $d \vec{x} = dX d\mu d\nu$, and $\delta(x)$ denotes the Dirac delta function, in the case of an operator argument being treated as $\delta{(\hat{A})} = (2\pi)^{-1} \int dk e^{i k \hat{A}}$. The purpose of these operators is to map operators used in the traditional formalism of quantum mechanics to $C$-functions
\begin{equation}
	f_{\hat{A}}(\vec{x}) = \operatorname{Tr}\{ \hat{A} \hat{\mathcal{U}}(\vec{x}) \}
	\label{eq:operator_to_symbol}
\end{equation}
referred to as operator symbols and, inversely,
\begin{equation}
	\hat{A} = \int f_{\hat{A}}(\vec{x}) \hat{\mathcal{D}}(\vec{x}) d\vec{x}.
	\label{eq:symbol_to_operator}
\end{equation}
Note, that the relationship between the symbols associated with an operator and its hermitian conjugate
\begin{equation}
	f_{\hat{A}^\dag}(\vec{x}) = f_{\hat{A}}^*(\vec{x})
	\label{eq:h.c.operator_symbols}
\end{equation}
immediately stems from conjugating \eqref{eq:operator_to_symbol}. The symbol of a multiplication of two operators, $\hat{A} \hat{B}$, is easily derived from \eqref{eq:operator_to_symbol} and \eqref{eq:symbol_to_operator}:
\begin{equation}
	f_{\hat{A} \hat{B}}(\vec{x}) = \iint f_{\hat{A}}(\vec{x}_1) f_{\hat{B}}(\vec{x}_2) \operatorname{Tr} \{ \hat{\mathcal{U}}(\vec{x}) \hat{\mathcal{D}}(\vec{x}_1) \hat{\mathcal{D}}(\vec{x}_2) \} d\vec{x}_1 d\vec{x}_2.
	\label{eq:ABproduct} 
\end{equation}
This relation, which is also convenient to shortly denote by
\begin{equation}
	f_{\hat{A} \hat{B}}(\vec{x}) = f_{\hat{A}}(\vec{x}) \star f_{\hat{B}}(\vec{x}),
	\label{eq:star_product}
\end{equation}
is called the star product of operator symbols. It is the symbol product that possess the operator's non-commutativity and hence brings this feature to the formalism of symplectic tomography.

Using the above mapping, we define a symplectic tomogram just as the symbol of the density matrix
\begin{equation}
	\mathcal{T}(\vec{x}) = \operatorname{Tr} \{ \hat{\rho} \hat{\mathcal{U}}(\vec{x}) \}.
	\label{eq:Tomogram_definition}
\end{equation}
It is rather easy to show, that a tomogram defined in this way is a real probability distribution, and a set of tomograms for all values of $\mu$ and $\nu$ characterizes the quantum state \cite{Ibort2009}. Having such a set, the density matrix is easily restored by the inverse transformation using dequantizer~\eqref{eq:UD}:
\begin{equation}
	\hat{\rho} = \int \mathcal{T}(\vec{x}) \hat{\mathcal{D}}(\vec{x}) d\vec{x}.
	\label{eq:inverse_transformation}
\end{equation}

The matrix elements of the quantizer operator can be obtained by a direct calculation. In result, in the coordinate representation these elements are
\begin{equation}
	\mathcal{D}_{qq'}(\vec{x}) \equiv \bra{q} \hat{\mathcal{D}}(\vec{x}) \ket{q'} = \frac{1}{2\pi} \exp \left[ i X - i \mu \frac{q + q'}{2} \right] \delta{(q - q' - \nu)},
	\label{eq:Dqq-prim}
\end{equation}
while in the discrete basis $\ket{m}$ these are
\begin{equation}
	\mathcal{D}_{nm}(\vec{x}) \equiv \bra{n} \hat{\mathcal{D}}(\vec{x}) \ket{m} = \frac{1}{2\pi} \exp \left[ i \left(X + \frac{\mu \nu}{2} \right) \right] \int \psi_n^*(q) \psi_m(q - \nu) e^{-i \mu q} dq.
	\label{eq:D_discrete_matrix_elements}
\end{equation}
In the particular case when the subscripts coincide, the matrix elements are expressed through the basis tomograms:
\begin{equation}
	\mathcal{D}_{mm}(\vec{x}) = \frac{1}{2\pi} \int \mathcal{T}_{m} (X + X', \mu, \nu) e^{-i X'} dX',
\end{equation}
where $\mathcal{T}_{m}(\vec{x}) \equiv \operatorname{Tr} \{ \ket{m}\bra{m} \hat{\mathcal{U}}(\vec{x}) \} = \mathcal{U}_{mm}(\vec{x})$ are the tomograms of the pure basis state $\ket{m}\bra{m}$.

Regarding the matrix elements of the dequantizer operator $\hat{\mathcal{U}} (\vec{x})$, they are nothing but the symbols of the operators $\ket{q'}\bra{q}$ in the continuous-variable coordinate basis or the operators $\ket{m}\bra{n}$ in the discrete basis. Their explicit expressions can easily be derived from \eqref{eq:Dqq-prim} and \eqref{eq:D_discrete_matrix_elements} if we notice that $\hat{\mathcal{U}}(\vec{x}) = \int \hat{\mathcal{D}}(k\vec{x}) dk$, which directly stems from \eqref{eq:UD}. In the coordinate representation this calculation results in
\begin{equation}
		\mathcal{U}_{qq'}(\vec{x}) = \begin{cases} \delta(q - q') \delta(X - \mu q), & \nu = 0, \\ \displaystyle \frac{1}{2\pi |\nu|} \exp \left[i (q - q') \frac{X}{\nu} - i (q^2 - q'^2) 	\frac{\mu}{2\nu} \right], & \nu \ne 0. \end{cases}
		\label{eq:Uqq-prim}
\end{equation}
In the discrete basis it is possible to express $\mathcal{U}_{nm}(\vec{x}) = \operatorname{Tr} \{ \ket{n}\bra{m} \hat{\mathcal{U}}(\vec{x}) \}$ through tomograms using the decomposition $\ket{m}\bra{n} = \hat{\rho}_{mn}^d + i \hat{\rho}_{mn}^r - 2^{-1} (1 + i) (\hat{\rho}_m + \hat{\rho}_n)$ where $\hat{\rho}_{mn}^d = 2^{-1} (\ket{m} + \ket{n})(\bra{m} + \bra{n})$, $\hat{\rho}_{mn}^r = 2^{-1} (\ket{m} + i\ket{n})(\bra{m} - i\bra{n})$, $\hat{\rho}_m = \ket{m}\bra{m}$ and $\hat{\rho}_n = \ket{n}\bra{n}$ indeed represent real quantum states and hence are expressible through tomograms.

It is convenient to define a scalar product of operator symbols associated with operators $\hat{A}$ and $\hat{B}$ as
\begin{equation}
	(f_{\hat{A}}, f_{\hat{B}}) = \frac{1}{2\pi} \int f_{\hat{A}}^*(X, \mu, \nu) f_{\hat{B}}(\bar{X}, \mu, \nu) e^{i (X - \bar{X})} d\bar{X} dX d\mu d\nu,
	\label{eq:symbols_product}
\end{equation}
that emerges naturally from the Hilbert-Schmidt product $\operatorname{Tr}\{ \hat{A}^\dag \hat{B} \}$ used in traditional quantum formalism. This product allows us to reformulate several familiar quantities in terms of symplectic tomograms. Of them, the basic and important one is the tomogram decomposition. Let the general state $\hat{\rho}$ be decomposed in the density matrix basis $\hat{\rho}_k$ as $\hat{\rho} = \sum_k \alpha_k \hat{\rho}_k$ where the decomposition coefficients are $\alpha_k = \operatorname{Tr} \{ \hat{\rho} \hat{\rho}_k \}$ and make sense of the probabilities the system jumps from the initial state $\hat{\rho}$ to the state $\hat{\rho}_k$. Due to linearity, the same is also valid in the tomogram representation of quantum states:
\begin{equation}
	\mathcal{T} = \sum_k \alpha_k \mathcal{T}_k,
\end{equation}
where the same decomposition coefficients can be shortly written in terms of the symbols product \eqref{eq:symbols_product} as $\alpha_k = (\mathcal{T}, \mathcal{T}_k)$. Another worth noting application of product \eqref{eq:symbols_product} is to multiply the tomogram by itself to indicate whether the state is mixed: $(\mathcal{T}, \mathcal{T})<1$ for a mixed state and $(\mathcal{T}, \mathcal{T}) = 1$ for a pure one.

\section{Quantum process in tomographic picture}

The operator-sum representation of a quantum process is a convenient and important formalism revealing the structure of the process and is undoubtedly useful when the behaviour of an open quantum system is of interest. To build this construction, we consider the system as a part of a larger bipartite system undergoing a joint unitary evolution. Then, the desired decomposition of the process emerges when tracing out the environment after the evolution. Transferring this formalism to the symplectic tomography terms is attractive since there is active fundamental and practical research in this area \cite{DAriano2002}. In this section, we will discuss how the operator-sum representation of a quantum process extends to the probability representation of quantum mechanics.

To be concrete, consider a bipartite quantum system and its composite Hilbert space $\mathbb{H} \otimes \mathbb{H}^E$, consisting of spaces associated with the system under consideration and the environment (the superscript $E$ denotes hereafter the part related to the environment). Let the system and environment initially be in the uncorrelated states $\hat{\rho}$ and $\hat{\rho}^E$, the latter for simplicity has been chosen to be diagonal in coordinate basis $\hat{\rho}^E = \int p(q) \ket{q} \bra{q} dq$. Having undergone a common unitary evolution, which is governed by the evolution operator $\hat{U}$ acting on both the system and environment, their joint state transforms as $\hat{\rho} \otimes \hat{\rho}^E \rightarrow \hat{U} \hat{\rho} \otimes \hat{\rho}^E \hat{U}^\dag$. If we now trace out the environment, we come to the relation
\begin{equation}
	\hat{\rho}' = \operatorname{Tr}_E \{ \hat{U} \hat{\rho} \otimes \hat{\rho}^E \hat{U}^\dag \} = \int \hat{A}_{qq'} \hat{\rho} \hat{A}_{qq'}^\dag dq dq',
	\label{eq:QO_traditional}
\end{equation}
revealing how the initial state of only the system of interest is transformed. The operators $\hat{A}_{qq'} \equiv \sqrt{p(q')} \bra{q} \hat{U} \ket{q'}$ are the Kraus operators in the coordinate representation of the environment, they act only on the system and all of them together characterize a particular quantum process.

To translate the operator-sum representation into terms of symplectic tomography, first we need to turn from the evolution operator $\hat{U}$ to its operator symbol $f_{\hat{U}}(\vec{x}, \vec{y}) = \operatorname{Tr} \{ \hat{U} \hat{\mathcal{U}} (\vec{x}, \vec{y})\}$, where $\hat{\mathcal{U}}(\vec{x}, \vec{y}) = \hat{\mathcal{U}}(\vec{x}) \hat{\mathcal{U}}^E(\vec{y})$ is the two-mode dequantizer \cite{Manko2003}. So, if we take the above definition of the Kraus operator and express the evolution operator standing there through its symbol using the two-mode quantizer operator $\hat{\mathcal{D}}(\vec{x}, \vec{y}) = \hat{\mathcal{D}}(\vec{x}) \hat{\mathcal{D}}^E(\vec{y})$, the Kraus operator will get the form
\begin{equation}
	\hat{A}_{qq'} = \sqrt{p(q')} \int f_{\hat{U}}(\vec{x}, \vec{y}) \mathcal{D}_{qq'}^E(\vec{y}) \hat{\mathcal{D}}(\vec{x}) d\vec{x} d\vec{y},
	\label{eq:Ai}
\end{equation}
where the matrix elements $\mathcal{D}_{qq'}^{E}(\vec{y}) \equiv \bra{q} \hat{\mathcal{D}}^E(\vec{y}) \ket{q'}$ related to the environment are actually defined in \eqref{eq:Dqq-prim}. Now, from comparing \eqref{eq:Ai} with definition of operator decomposition \eqref{eq:symbol_to_operator}, it becomes immediately apparent that the symbol of the Kraus operator is
\begin{equation}
	f_{\hat{A}_{qq'}}(\vec{x}) = \sqrt{p(q')} \int f_{\hat{U}}(\vec{x}, \vec{y}) \mathcal{D}_{qq'}^{E}(\vec{y}) d\vec{y}.
	\label{eq:Kraus_operator_symbol}
\end{equation}

Finally, turning operator-sum representation \eqref{eq:QO_traditional} of a quantum process to the tomography picture, it takes the form
\begin{equation}
	\mathcal{T}'(\vec{x}) = \int f_{\hat{A}_{qq'}} \star \mathcal{T} \star f_{\hat{A}_{qq'}}^* dq dq' = \int \mathcal{T}(\vec{\bar{x}}) K_{qq'}(\vec{\bar{x}}, \vec{x}) d\vec{\bar{x}} dq dq',
	\label{eq:QO_in_PR}
\end{equation}
where $\mathcal{T}(\vec{x})$ and $\mathcal{T}'(\vec{x})$ represent the initial and final states of the system respectively, and the kernels $K_{qq'}(\vec{\bar{x}}, \vec{x})$, hereafter just called partial quantum operations, are
\begin{equation}
	K_{qq'}(\vec{\bar{x}}, \vec{x}) = \int f_{\hat{A}_{qq'}}(\vec{x}_1) f_{\hat{A}_{qq'}}^* (\vec{x}_2) \operatorname{Tr}\{ \hat{\mathcal{D}}(\vec{x}_1) \hat{\mathcal{D}}(\vec{\bar{x}}) \hat{\mathcal{D}}(\vec{x}_2) \hat{\mathcal{U}}(\vec{x}) \} d\vec{x}_1 d\vec{x}_2.
	\label{eq:Ki_definition}
\end{equation}
The trace this partial operation incorporates is essentially that used to derive the operator symbol of multiplication of three operators, and its direct calculation gives
\begin{equation}
	\begin{aligned}
		&\operatorname{Tr}\{ \hat{\mathcal{D}}(\vec{x}_1) \hat{\mathcal{D}}(\vec{\bar{x}}) \hat{\mathcal{D}}(\vec{x}_2) \hat{\mathcal{U}}(\vec{x}) \} = \frac{1}{(2\pi)^3} \exp \left[ i \left( \bar{X} + X_1 + X_2 - X \frac{\bar{\nu} + \nu_1 + \nu_2}{\nu} \right) \right] \\
		& \cross \exp \left[ \frac{i}{2} \Big( \bar{\mu} (\nu_1 - \nu_2) - (\mu_1 - \mu_2) \bar{\nu} - \mu_1 \nu_2 + \mu_2 \nu_1 \Big) \right] \delta((\bar{\mu} + \mu_1 + \mu_2) \nu - (\bar{\nu} + \nu_1 + \nu_2) \mu). \\
	\end{aligned}
\end{equation}

Now let us go back to relationship \eqref{eq:QO_in_PR}. Performing there a preliminary integration over the coordinate, we get the kernel of the overall quantum process
\begin{equation}
	\tilde{K}(\vec{\bar{x}}, \vec{x}) = \int K_{qq'}(\vec{\bar{x}}, \vec{x}) dq dq'. \\
\end{equation}
The state transformation $\mathcal{T}'(\vec{x}) = \int \mathcal{T}(\vec{\bar{x}}) \tilde{K}(\vec{\bar{x}}, \vec{x}) d\vec{\bar{x}}$ involved by the process resembles that of a "classical" propagator (tomogram propagator) \cite{Mancini1997}. Indeed, it becomes exactly a tomogram propagator whenever the quantum process reduces to unitary.

As in traditional formalism, where a set of operators must satisfy the completeness condition in order to represent a quantum process, there is a similar rule in the probability representation. Namely, this condition is imposed to a set of operator symbols $f_{a}(\vec{x})$ and reads
\begin{equation}
	\begin{aligned}
		\frac{1}{(2\pi)^2} \int f_{a}(X, \mu, \nu) &f_{a}^* (X', \mu + \bar{\mu}, \nu + \bar{\nu})  \\
		& \cross \exp \left[ i \left( X - X' + \frac{\bar{\mu} \nu - \mu \bar{\nu}}{2} \right) \right] dX dX' d\mu d\nu da = \delta(\bar{\mu}) \delta(\bar{\nu}). \\
	\end{aligned}
	\label{eq:completeness_condition}
\end{equation}
For derivation of this relationship, refer to \ref{A_complenetess_derivation}.

Whenever it is convenient to employ a discrete basis, the above formulas can be easily adapted for that. In essence, the discrete case is distinguished almost just by turning from the integration over coordinates to summing over indices $m$ and $n$. Indeed, now the Kraus operators take the form $\hat{A}_{mn} \equiv \sqrt{p_n} \bra{m} \hat{U} \ket{n}$, hence their symbols are
\begin{equation}
	f_{\hat{A}_{mn}} = \sqrt{p_n} \int f_{\hat{U}}(\vec{x}, \vec{y}) \mathcal{D}_{mn}^E(\vec{y}) d\vec{y},
\end{equation}
where the matrix elements $\mathcal{D}_{mn}^E(\vec{x})$ are defined in \eqref{eq:D_discrete_matrix_elements} and associated with the environment. Then, given the Kraus operators in the matrix representation, the partial quantum operation in the tomographic picture takes the form
\begin{equation}
	K_{mn} (\vec{\bar{x}}, \vec{x}) = \sum\limits_{ijkl} A_{mn,ij} A_{mn,lk}^* \mathcal{D}_{jk}(\vec{\bar{x}}) \mathcal{U}_{li}(\vec{x}).
	\label{eq:Ki_discrete}
\end{equation}

A quantum process can be expressed in a different way employing scalar product \eqref{eq:symbols_product} of operator symbols. To illustrate this, recall the general relation for the tomogram of the transformed system $\mathcal{T}'(\vec{x}) = \sum_{mn} \operatorname{Tr}\{ \hat{A}_{mn} \hat{\rho} \hat{A}_{mn}^\dag \hat{\mathcal{U}}(\vec{x}) \}$ where, for definiteness, we use the discrete basis. Using the possibility of cyclic permutation of operators under the trace, the same can be written as 
\begin{equation}
	\mathcal{T}'(\vec{x}) = \sum_{mn} \operatorname{Tr}\{ \hat{\rho} \hat{\mathcal{U}}_{mn}(\vec{x}) \},
	\label{eq:tomogram_using_Umn}
\end{equation}
where $\hat{\mathcal{U}}_{mn} \equiv \hat{A}_{mn}^\dag \hat{\mathcal{U}} \hat{A}_{mn}$. From comparing \eqref{eq:tomogram_using_Umn} with \eqref{eq:Tomogram_definition} we see, that the sum $\sum_{mn} \hat{\mathcal{U}}_{mn}$ play here the role of the dequantizer after the operation is performed. Relationship \eqref{eq:tomogram_using_Umn}, in turn, being written in terms of the scalar product of symbols becomes
\begin{equation}
	\mathcal{T}'(\vec{x}) = \sum_{mn} ( \mathcal{T}(\vec{\bar{x}}), f_{\hat{\mathcal{U}}_{mn}(\vec{x})} (\vec{\bar{x}}) ),
\end{equation}
where $f_{\hat{\mathcal{U}}_{mn}(\vec{x})} (\vec{\bar{x}}) = f_{\hat{A}_{mn}}^*(\vec{\bar{x}}) \star f_{\hat{\mathcal{U}}(\vec{x})}(\vec{\bar{x}}) \star f_{\hat{A}_{mn}}(\vec{\bar{x}})$. Here $f_{\hat{\mathcal{U}}(\vec{x})}(\vec{\bar{x}})$ is the symbol of the dequantizer, the direct deriving of which leads to a formal expression
\begin{equation}
	f_{\hat{\mathcal{U}}(\vec{x})}(\vec{\bar{x}}) = \frac{1}{2\pi} \int  e^{i k \left( \bar{X} \mu - X \bar{\mu} \right)} \delta\left( k( \mu \bar{\nu} - \bar{\mu} \nu) \right) dk.
\end{equation}
The crucial point is that this integral is obviously diverges. However, note that this symbol always stands as a multiplier in the star product or scalar product of symbols, where it is integrated with $e^{-i X}$ over $X$ (see \eqref{eq:symbols_product} and an explicit expression for the star-product kernel, e.g., in \cite{Manko2021}). The result of such integration, instead, is a regular function:
\begin{equation}
	\int f_{\hat{\mathcal{U}}(\vec{\bar{x}})}(\vec{x}) e^{-i X} dX = \delta\left( \bar{\mu} \nu - \mu \bar{\nu} \right) e^{- i \bar{X} \mu/\bar{\mu} }.
\end{equation}
Therefore, the symbol of the dequantizer $f_{\hat{\mathcal{U}}(\vec{x})}(\vec{\bar{x}})$ should be treated in a generalized sense, similarly to a delta-function.

\section{Examples of processes in tomographic picture}

In this section, we will discuss several particular quantum processes regarding them in the probability representation of quantum mechanics.

\subsection{Von Neumann measurement model}

In some applications, it may be attractive to realize the von Neumann measurement model, being one of the basic practical procedures to find out in what state the system was  \cite{Das2014, Das2017}. The essence of such a measurement is to bring the system under investigation to interaction with a measuring device (pointer), whose indication can be retrieved classically. In particular, by lowing the interaction strength, one turns to the weak measurement regime, the latter nowadays is the subject being actively developed \cite{Gross2015}. To refine the pointer state in this procedure, one can apply the balanced homodyning technique \cite{DAriano1997}. Therefore, reformulation of the measuring process in terms of the probability representation of quantum mechanics could help in experimental data processing or new methods designing.

First, let us outline the von Neumann measurement model as it is. Consider the quantum system and its observable $\hat{A}$ of interest, whose eigenvectors are $\ket{a_i}$ with corresponding eigenvalues $a_i$, $i = 1 \dots N$. Let the system be in the state $\ket{\psi_0} = \sum_{i=1}^N c_i \ket{a_i}$. The measuring device is assumed to have continuous indications, so it can be described by an infinite-dimensional Hilbert space with conjugate coordinate and moment operators $\hat{q}$ and $\hat{p}$ acting in this space. Since we want the pointer to also contain the classical character, its initial state is taken to be Gaussian $\ket{\phi_0} = \left( \kappa/2\pi\right)^{1/4} \int e^{- \kappa q^2/4} \ket{q} dq$.

Following the von Neumann basic model, the system-pointer interaction is instantaneous and is described by interaction Hamiltonian $\hat{H}_{int}(t) = g \delta(t - t_0) \hat{A} \otimes \hat{p}$, where $g$ is the coupling strength. Using it, the evolution operator takes the simple form 
\begin{equation}
	\hat{U} = e^{-i \int \hat{H}_{int} dt} = e^{-i g \hat{A} \otimes \hat{p}}.
	\label{eq:U_vonNeumann_measurement}
\end{equation}
The final system-pointer state is then becomes
\begin{equation}
	\hat{U} \ket{\psi_0} \otimes \ket{\phi_0} = \left( \frac{\kappa}{2\pi} \right)^{1/4} \sum_{i=1}^N \int c_i e^{- \kappa (q - g a_i)^2/4} \ket{a_i} \otimes \ket{q} dq,
\end{equation}
the pointer part of which is composed of Gaussians centered in $g a_i$. It is these peaks that allow determining the probability that the system initially could be found in the state $i$, provided that they are separated enough (the strong coupling regime). On the contrary, if we turn $g$ to be smaller until peaks substantially overlap, we invoke the weak measurement regime. Tracing out non-interesting part, one gets the Kraus operators in the operator-sum representation of the process
\begin{equation}
	\hat{M}_Q \equiv \bra{Q} \hat{U} \ket{\phi_0} = \left( \frac{\kappa}{2\pi} \right)^{1/4} e^{- \kappa (Q - g \hat{A})^2/4}
	\label{eq:MQ_Kraus_operator}
\end{equation}
if we observe only the system and
\begin{equation}
	\hat{L}_j \equiv \bra{a_j} \hat{U} \ket{\psi_0} = c_j e^{-i g a_j \hat{p}}
	\label{eq:Lj_Kraus_operator}
\end{equation}
if we are interested just in the pointer state.

Now let's reformulate the above process in the terms of the probability representation. To follow prescription \eqref{eq:Kraus_operator_symbol} of deriving the symbols of the Kraus operators, we are first needed to find the symbol associated with evolution operator \eqref{eq:U_vonNeumann_measurement}. The direct calculation of it gives
\begin{equation}
	f_{\hat{U}} (\vec{x}, \vec{y}) = \sum_i \int \delta(q' - q - g a_i) \mathcal{U}_{ii} (\vec{x}) \mathcal{U}_{qq'}^E (\vec{y}) dq dq'. \\
\end{equation}
Using it in \eqref{eq:Kraus_operator_symbol} then yields the symbols
\begin{equation}
	\begin{aligned}
		& f_{\hat{M}_Q} (\vec{x}) = \left( \frac{\kappa}{2\pi} \right)^{1/4} \sum_i e^{- \kappa (Q - g a_i)^2/4} \mathcal{U}_{ii} (\vec{x}), \\
		& f_{\hat{L}_j} (\vec{x}) = c_j \int e^{i kX} \delta(k\mu) \delta{(g a_j + k\nu)} dk \\
	\end{aligned}
\end{equation}
for Kraus operators \eqref{eq:MQ_Kraus_operator} and \eqref{eq:Lj_Kraus_operator} respectively, here $\mathcal{U}_{ii} \equiv \bra{a_i} \hat{\mathcal{U}}_{ii} \ket{a_i} = \mathcal{T}_i$ is  the tomogram of the system eigenstate. Substituting now these symbols to \eqref{eq:Ki_definition}, it is straightforward to get the kernels of the partial operations:
\begin{equation}
	\begin{aligned}
		K^{(S)}_{Q}(\vec{\bar{x}}, \vec{x}) = \left( \frac{\kappa}{2\pi} \right)^{1/2} &\sum_{i,j} e^{- \kappa (Q - g a_i)^2/4 - \kappa (Q - g a_j)^2/4} \\
		& \cross \int \mathcal{T}_{i} (\vec{x}_1) \mathcal{T}_{j}^* (\vec{x}_2) \operatorname{Tr}\{ \hat{\mathcal{D}}(\vec{x}_1) \hat{\mathcal{D}}(\vec{\bar{x}}) \hat{\mathcal{D}}(\vec{x}_2) \hat{\mathcal{U}}(\vec{x}) \} d\vec{x}_1 d\vec{x}_2 \\
	\end{aligned}
\end{equation}
from the system viewpoint and
\begin{equation}
	K_{j}^{(P)}(\vec{\bar{x}}, \vec{x}) = \frac{|c_j|^2}{2\pi} \exp \left[ i \left( \bar{X} - X \frac{\bar{\nu}}{\nu} + \bar{\mu} g a_j \right) \right] \delta(\bar{\mu} \nu - \bar{\nu} \mu) \\
\end{equation}
for the pointer. Integrating (summing) them, we get the full kernels of the respective operations:
\begin{equation}
	\begin{aligned}
		\tilde{K}^{(S)}(\vec{\bar{x}}, \vec{x}) = & \sum_{i,j} e^{-\frac{\kappa g^2 (a_i - a_j)^2}{8}} \int \mathcal{T}_{i} (\vec{x}_1) \mathcal{T}_{j}^* (\vec{x}_2) \operatorname{Tr}\{ \hat{\mathcal{D}}(\vec{x}_1) \hat{\mathcal{D}}(\vec{\bar{x}}) \hat{\mathcal{D}}(\vec{x}_2) \hat{\mathcal{U}}(\vec{x}) \} d\vec{x}_1 d\vec{x}_2 \\
	\end{aligned}
\end{equation}
and
\begin{equation}
	\tilde{K}^{(P)}(\vec{\bar{x}}, \vec{x}) = \sum_j \frac{|c_j|^2}{2\pi} \exp \left[ i \left( \bar{X} - X \frac{\bar{\nu}}{\nu} + \bar{\mu} g a_j \right) \right] \delta(\bar{\mu} \nu - \bar{\nu} \mu). \\
\end{equation}

\subsection{Projective measurement}

A significant type of the system-environment interaction is the projective measurement carried out over the system. However, in real experiment, we always have limited precision, regardless of whether this is due to imperfect measuring device or the precision was deliberately reduced to minimize the back action (weak measurement) \cite{Hofmann2010}. Therefore, the formulation of this kind of measurement within the probability representation of quantum mechanics is of interest, in particular, for quantum state engineering or process control.

Let's derive the process kernel describing a projective measurement. Of these, the basic one is the projection to a state with a certain coordinate $a$. The Kraus operators associated with such a process are
\begin{equation}
	\hat{\Pi}_a = \ket{a} \bra{a}
\end{equation}
for all $a$. The symbols of these operators are
\begin{equation}
	f_{\hat{\Pi}_a}(\vec{x}) \equiv \mathcal{U}_{aa}(\vec{x}) = \frac{1}{2\pi} \int e^{ik (X - \mu a)} \delta(k\nu) dk,
	\label{eq:Uaa}
\end{equation}
which diverge, so should be treated in a generalized sense (refer to the remark concluding the preceding section). Using \eqref{eq:Ki_definition}, one gets the kernel of the partial operation
\begin{equation}
	K_a(\vec{\bar{x}}, \vec{x}) = \mathcal{D}_{aa}(\vec{\bar{x}}) \mathcal{U}_{aa}(\vec{x}),
\end{equation}
which actually describes a selective coordinate measurement giving the outcome $a$, and also should be treated in a generalized sense. The function $\mathcal{D}_{aa}(\vec{\bar{x}})$ here is the diagonal element of the quantizer operator, whose explicit expression follows from \eqref{eq:Dqq-prim}. Integrating this kernel with an arbitrary initial state $\mathcal{T}(\vec{x})$ yields the final state
\begin{equation}
	\mathcal{T}_a'(\vec{x}) = \frac{1}{2\pi} \mathcal{U}_{aa}(\vec{x}) \int \mathcal{T}(\bar{X}, \bar{\mu}, 0) e^{i (\bar{X} - \bar{\mu} a)} d\bar{X} d\bar{\mu} \\ 
\end{equation}
the system gets after measuring its coordinate, provided that the outcome is $a$. Note, that $\mathcal{T}_a'(\vec{x})$ is not a true tomogram, but is the tomogram density in the set of outcomes $\{ a \}$ and represents the result of a selective measurement.

To obtain the kernel associated with a non-selective measurement of the system position, the kernels of all partial operations must be integrated over the set of outcomes $\tilde{K}(\vec{\bar{x}}, \vec{x}) = \int K_a(\vec{\bar{x}}, \vec{x}) da$, that results in
\begin{equation}
	\tilde{K}(\vec{\bar{x}}, \vec{x}) = \frac{1}{2\pi} e^{i (\bar{X}-X \bar{\mu}/\mu)} \delta(\bar{\nu}) \delta(\bar{\mu} \nu).
\end{equation}
Then, as it follows from \eqref{eq:QO_in_PR}, the projection to the state with a certain coordinate gives the tomogram
\begin{equation}
	\mathcal{T}'(\vec{x}) = \int \mathcal{T}(a, 1, 0) \mathcal{U}_{aa}(\bar{x}) da.
	\label{eq:Tfinal_selective_certain-coord}
\end{equation}
The first multiplier under the integral is nothing but the coordinate distribution of the initial state: $\mathcal{T}(a, 1, 0) = \bra{a} \hat{\rho} \ket{a}$. Hence, final tomogram \eqref{eq:Tfinal_selective_certain-coord} rewritten in this fashion, is just $\mathcal{T}'(\vec{x}) = \int \bra{a} \hat{\rho} \ket{a} \bra{a} \hat{\mathcal{U}}(\vec{x}) \ket{a} da$. This result is actually expected if we choose a straightforward way of finding the final tomogram from the density matrix $\int \hat{\Pi}_a \hat{\rho} \hat{\Pi}_a da$ which the system obtain after projecting its state to that with a certain coordinate.

The above consideration can be naturally generalized to a Gauss-type projection measurement of position, which is described by a set of Kraus operators
\begin{equation}
	\hat{\Pi}_a = \left( \frac{1}{\pi \kappa^2} \right)^{-1/4} \int \exp \left[ - \frac{(q - a)^2}{2\kappa^2} \right] \ket{q} \bra{q} dq
	\label{eq:Pi_gaussian}
\end{equation}
for all $a$. A direct calculation of the symbols corresponding to these operators gives
\begin{equation}
	f_{\hat{\Pi}_a}(\vec{x}) = \left( \frac{\kappa^2}{4\pi^3} \right)^{1/4} \int \exp \left[ ik(X - \mu a) - \frac{\kappa^2 k^2 \mu^2}{2} \right] \delta(k\nu) dk,
\end{equation}
which are also treated in a generalized sense. Then the kernel of selective position measurement, corresponding to the outcome $a$, is got using \eqref{eq:Ki_definition}:
\begin{equation}
	K_a(\vec{\bar{x}}, \vec{x}) =
 		\begin{cases}
			\displaystyle \frac{1}{2\pi} \delta(X) \delta(\bar{\nu}) \exp \left[i(\bar{X} - \bar{\mu} a) - \frac{\kappa^2 \bar{\mu}^2}{4} \right], & \nu = 0,~\mu = 0, \\~\\
			\displaystyle \frac{1}{2 \pi \sqrt{\pi} |\mu|} \delta(\bar{\nu}) \exp \left[ i \frac{\bar{X} \mu - X \bar{\mu}}{\mu} - \frac{(X - \mu a)^2}{\kappa^2 \mu^2} \right], & \nu = 0,~\mu \ne 0, \\~\\
 			\displaystyle \frac{1}{4\pi^2 |\nu|}  \exp \left[ i\frac{\bar{X} \nu - X \bar{\nu} - (\bar{\mu} \nu - \mu \bar{\nu})a}{\nu} - \frac{\kappa^2 (\bar{\mu}\nu - \mu\bar{\nu})^2 + \kappa^{-2} \bar{\nu}^2 \nu^2}{4\nu^2}\right], & \nu \ne 0.
			\end{cases}
	\label{eq:Ka_selective_Gauss}
\end{equation}
This kernel leads to the final tomogram density
\begin{equation}
	\begin{aligned}
		\mathcal{T}_a'(\vec{x}) = \frac{1}{(2\pi)^2 |\nu|} \int &\mathcal{T}(\vec{\bar{x}}) \exp \left[ i\frac{\bar{X} \nu - X \bar{\nu} - (\bar{\mu} \nu - \mu \bar{\nu})a}{\nu} \right] \\
		& \cross \exp \left[ - \frac{\kappa^2 (\bar{\mu}\nu - \mu\bar{\nu})^2 + \kappa^{-2} \bar{\nu}^2 \nu^2}{4\nu^2} \right] d\vec{\bar{x}} \\ 
	\end{aligned}
	\label{eq:Tfinal_selective_Gauss}
\end{equation}
the system obtains after the Gaussian-type projection of the coordinate. Integrating kernel \eqref{eq:Ka_selective_Gauss} and final tomogram density \eqref{eq:Tfinal_selective_Gauss} over outcome set, we get, respectively, their counterparts for a non-selective position measurement:
\begin{equation}
	\tilde{K}(\vec{\bar{x}}, \vec{x}) = \frac{1}{2\pi} \exp \left[ i \frac{\bar{X} \mu - X \bar{\mu}}{\mu} - \frac{\bar{\nu}^2}{4 \kappa^2} \right] \delta(\bar{\mu} \nu - \mu \bar{\nu})
\end{equation}
and
\begin{equation}
	\mathcal{T}'(\vec{x}) = \left( \frac{\kappa^2}{\pi \nu^2} \right)^{1/2} \int \mathcal{T}(\bar{X}, \mu , \nu) \exp \left[ - \frac{\kappa^2(\bar{X} - X)^2}{\nu^2} \right] d\bar{X}. 
\end{equation}
We see that the measurement of this type leads to a Gaussian-type blur of the tomogram dependence on $X$. Note that in contrast to the preceding case, when the system state was projected onto that with a certain coordinate, a Gauss-type projection yields a tomogram that is expressed by a regular function.

If the discrete basis is more convenient to use, the process kernel associated with a projective measurement can be derived similarly. In particular, the matrix projecting an arbitrary initial state to the $m$-th basis state is $A_{m,ij} = \delta_{ij} \delta_{jm}$, where $\delta_{ij}$ is the Kronecker symbol. Calculating the kernels of partial operations using \eqref{eq:Ki_discrete} leads to 
\begin{equation}
	K_m(\vec{\bar{x}}, \vec{x}) = \mathcal{D}_{mm}(\vec{\bar{x}}) \mathcal{U}_{mm}(\vec{x}) = \frac{1}{2\pi} \mathcal{T}_{m} (\vec{x}) \int \mathcal{T}_{m} (\bar{X} + X', \bar{\mu}, \bar{\nu}) e^{-i X'} dX',
\end{equation}
where $\mathcal{T}_{m}(\vec{x}) \equiv \operatorname{Tr}\{ \ket{m} \bra{m} \mathcal{U} (\vec{x})\}$ are the tomograms of the $m$-th basis state. Generalization of this projection to a Gauss-type one can be done using the Kraus operators $\hat{\Pi}_a = N^{-1} \sum_n \exp[-\kappa^2(n - a)^2/4] \ket{n} \bra{n}$, whose symbols are
\begin{equation}
	f_{\hat{\Pi}_a}(\vec{x}) = \frac{1}{N} \sum_n \exp \left[ -\frac{\kappa^2(n - a)^2}{4} \right] \mathcal{T}_{n}(\vec{x}),
\end{equation}
where $N$ is the normalizing coefficient. Substituting directly the Kraus operators to \eqref{eq:Ki_discrete}, one come to the relationship for the kernel of projecting to a certain basis state
\begin{equation}
	\begin{aligned}
		K_{a} (\vec{\bar{x}}, \vec{x}) &= \frac{1}{N^2} \sum\limits_{nm} \exp \left[-\frac{\kappa^2[(n - a)^2 + (m - a)^2]}{4} \right] \mathcal{D}_{nm}(\vec{\bar{x}}) \mathcal{U}_{mn}(\vec{x}).
	\end{aligned}
\end{equation}
Somewhat tedious, but straightforward derivation of the last terms gives that they are expressed through the basis tomograms as
\begin{equation}
	\begin{aligned}
		\mathcal{D}_{nm}(\vec{\bar{x}}) \mathcal{U}_{mn}(\vec{x}) = \frac{1}{(2\pi)^3} &\int \mathcal{T}_n(X' - \bar{X}, \mu' - \bar{\mu}, \nu' - \bar{\nu}) \mathcal{T}_m(X'' + kX, \mu' + k\mu, \nu' + k\nu) \\
		& \cross \exp \left\{ i \left[X' - X'' + \frac{\bar{\mu} \nu' - \mu' \bar{\nu}}{2} + k \frac{\mu' \nu - \mu \nu'}{2} \right] \right\} d\vec{x}' dX'' dk. \\
	\end{aligned}
\end{equation}

\subsection{Continuous measurement}

The special case of a quantum process concerns continuous measurement. The latter composes a rapidly developing area of quantum research, aimed not only at measuring purposes \cite{Silberfarb2005} but also at controlling quantum systems \cite{Wu2007}. The continuous measurement formalism within the tomographic representation is investigated in detail in Ref. \cite{Przhiyalkovskiy2021} and uses the restricted path integral approach, being especially intuitive and promising to study continuous decoherence processes \cite{Mensky2003, Mensky1994}. Moreover, it is favourably distinguished by its suitability to study spectral measurements when one is interested in particular spectral components of the observable.

As the formalism of symplectic tomography prescripts, one way to get the final tomogram of the isolated system, which evolves unitarily, is to convolve its initial tomogram with a tomogram propagator \cite{Mancini1997}. The latter is a function that in a certain sense is a counterpart of the quantum propagator used in conventional formalism of quantum mechanics. If the system undergoes a continuous measurement, an additional dissipation occurs. Its incorporation into the tomogram propagator formalism yields a set of partial propagators, which actually corresponds to the set of partial operations \eqref{eq:Ki_definition}. More specifically, it is the partial tomogram propagator $\Pi_{t,a}(\vec{\bar{x}}, \vec{x})$ relevant to a particular measurement outcome $a(t)$ that now plays the role of the partial operation kernel:
\begin{equation}
	\begin{aligned}
		K_a(\vec{\bar{x}}, \vec{x}) &= \Pi_{t,a}(\vec{\bar{x}}, \vec{x}) \equiv \frac{1}{4\pi^2} \int k^2 U_{T,a}^*(q_{f,1} + k\nu, q_{i,2}) U_{T,a}(q_{f,1}, q_{i,2} + k \nu') \\
		& \cross \exp \left[ik \left( X' + X - k \frac{\mu' \nu' + \mu \nu}{2} \right)  - i k( \mu' q_{i,2} + \mu q_{f,1} ) \right] dq_{f,1} dq_{i,2} dk,
	\end{aligned}
\end{equation}
where
\begin{equation}
	U_{T, a}(q_{f}, q_{i}) = \int\limits_{q_{i}, 0}^{q_{f}, T} d[q] w_a[q] e^{i S[q]}
\end{equation}
is the restricted path integral of the action $S[q]$ over paths $q(t)$, beginning in $q_{i}$ at time $t = 0$, ending in $q_{f}$ at time $t = T$, and which are weighted by a functional $w_a[q]$ according to the particular measurement under consideration. Among these, for example, the simplest is \cite{Calarco1995}
\begin{equation}
	w_a[q] = \exp \left\{ - \frac{2}{\Delta a^2 T} \int\limits_0^T (A(q, t) - a(t))^2 dt \right\},
	\label{eq:w_a}
\end{equation}
where $q(t)$ is the particular path, $a(t)$ is the measurement result and $\Delta a$ represents the measuring precision. As seen, this functional provides the lower the path weight, the farther the observable $A$ is from the measurement outcome at scale $\Delta a$. In particular, in the limit $T \rightarrow 0$ the measurement reduces to the impulsive one described by \eqref{eq:Pi_gaussian}. 

In the conclusion note that, in general, the restricted path integral formalism is capable to describe not only Markovian dynamics but also non-Markovian dynamics. Indeed, taking the weight functional of the Gaussian type as in \eqref{eq:w_a}, one can reduce the path integral approach to the differential form, namely, to a GKSL-type equation \cite{Mensky1994}, which approximates the Markovian case. On the other hand, giving the target observable in \eqref{eq:w_a} the form $\bar{A}(t) = \int P_t(t') A(t') dt'$ with appropriately chosen function $P_t(t')$ for a specific problem, it becomes able to describe the non-Markovian dynamics. In particular, it is then impossible to cast the evolution equation expressed through the restricted path integral to a time differential equation \cite{Mensky1997}. Therefore, this formalism can be used to cover non-Markovian dynamics of open quantum systems in the probability representation of quantum mechanics.

\subsection{Qubit operations}

Although the probability representation of quantum mechanics is most oriented to continuous-variable systems, the discrete case is also of interest. It concerns primarily the qubit state tomography, which is relevant for quantum computing, quantum cryptography, and the related fields \cite{Fedorov2013-3, Kiktenko2014}. In this subsection, we demonstrate how several basic quantum processes acting on a qubit look like from a tomographic standpoint. First, consider a process that flips the phase of a qubit with the probability $1-p$. Its Kraus operators written in the matrix form are
\begin{equation}
	\begin{aligned}
		& A_{1,nm} = \sqrt{p} \begin{pmatrix} 1 & 0 \\ 0 & 1 \end{pmatrix}, & A_{2,nm} = \sqrt{1 - p} \begin{pmatrix} 1 & 0 \\ 0 & -1 \end{pmatrix}. \\
	\end{aligned}
\end{equation}
The first matrix $A_{1,nm}$ obviously leaves the state unchanged with probability $p$. Hence, the kernel of this partial operation in the probability representation is just
\begin{equation}
	K_1(\vec{\bar{x}}, \vec{x}) = p \delta(\vec{\bar{x}} - \vec{x}).
\end{equation}
The derivation of the kernel induced by the matrix $A_{2,nm}$ yields
\begin{equation}
	\begin{aligned}
		K_2(\vec{\bar{x}}, \vec{x}) = (1 - p) &\bigg( \frac{1}{\pi} \mathcal{T}_{0} (\vec{x}) \int \mathcal{T}_{0} (\bar{X} + X', \bar{\mu}, \bar{\nu}) e^{-i X'} dX'  \\
		& + \frac{1}{\pi} \mathcal{T}_{1} (\vec{x}) \int \mathcal{T}_{1} (\bar{X} + X', \bar{\mu}, \bar{\nu}) e^{-i X'} dX' - \delta(\vec{\bar{x}} - \vec{x}) \bigg), \\
	\end{aligned}
\end{equation}
where $\mathcal{T}_{0}$ and $\mathcal{T}_{1}$ are the qubit basis states.

Another process to which we can easily give the tomographic form is amplitude damping. This process is characterized by the Kraus operators
\begin{equation}
	\begin{aligned}
		& A_{1} = \begin{pmatrix} 1 & 0 \\ 0 & \sqrt{1 - \gamma}	\end{pmatrix}, & A_{2} = \begin{pmatrix} 0 & \sqrt{\gamma} \\ 0 & 0	\end{pmatrix}. \\
	\end{aligned}
\end{equation}
Turning to the probability representation, the kernels corresponding to these partial operations are
\begin{equation}
	\begin{aligned}
		K_1&(\vec{\bar{x}}, \vec{x}) = \frac{\gamma}{2\pi} \mathcal{T}_{0} (\vec{x}) \int \mathcal{T}_{1} (\bar{X} + X_1, \bar{\mu}, \bar{\nu}) e^{-i X_1} dX_1 \\
		K_2&(\vec{\bar{x}}, \vec{x}) = \frac{1 - \sqrt{1 - \gamma}}{2\pi} \mathcal{T}_{0} (\vec{x}) \int \mathcal{T}_{0} (\bar{X} + X', \bar{\mu}, \bar{\nu}) e^{-i X'} dX' \\
		& + \frac{1 - \gamma - \sqrt{1 - \gamma}}{2\pi} \mathcal{T}_{1} (\vec{x}) \int \mathcal{T}_{1} (\bar{X} + X', \bar{\mu}, \bar{\nu}) e^{-i X'} dX' \\
		& + \sqrt{1 - \gamma} \delta(\vec{\bar{x}} - \vec{x}).
	\end{aligned}
\end{equation}

\section{Conclusion}

General formalism describing a quantum process is of undoubtedly great importance both theoretically and experimentally. In this work, the operator-sum representation of quantum processes, widely used within the traditional formalism of quantum mechanics, is extended to the probability representation of quantum mechanics. 

The main result of the work is a general relationship that expresses how a quantum process transforms a set of symplectic tomograms associated with a quantum state. It is exhibited, that this mapping, which connects the initial state with the final, is broken into partial operations in the same manner as the traditional formalism suggests by introducing the Kraus operators. These operators, in turn, appear in the probability representation in the form of their operator symbols, which comprise the symbol of the evolution operator associated with the joint unitary evolution of the investigated system and an ancillary environment. The constraint is derived to which a set of operator symbols must satisfy in order to represent the Kraus operators and therefore make up a quantum process.

To illustrate the operator-sum representation adapted to the probability representation of quantum mechanics, a number of specific examples were elaborated. Thus, we have considered the base von Neumann measurement model as well as obtained the kernel of both the projection of the system state onto a certain state and the fuzzy projection of the Gaussian type. The derived kernels of partial operations comprising the projection, in essence, describe a selective measurement and actually give a tomogram density in a set of measurement results, which are to be normalized by the appropriate probability in order to give a true tomogram. In fact, the expression for the tomogram density of a system undergone the projection turns out to diverge but may be treated in a generalized sense similar to a delta function. This peculiarity however can be worked around by using the Gaussian projection instead, which ensures tomograms to be regular. The kernel of the entire process, namely non-selective projective measurement, is obtained by integrating all kernels of the partial operations over the outcome set. In addition, such simplest processes acting on a qubit as its flipping and amplitude damping are also demonstrated.

\appendix

\section{The completeness condition for symbols of Kraus operators}
\label{A_complenetess_derivation}

Our aim here is to derive a completeness condition that operator symbols must obey to represent Kraus operators, namely a counterpart of the constraint $\sum \hat{K}_i^\dag \hat{K}_i = \hat{1}$ imposed on Kraus operators. Our starting point is the general equation governing the particular quantum process in the probability representation
\begin{equation}
	\mathcal{T}'(\vec{x}) = \int \mathcal{T}(\vec{\bar{x}}) K_{a}(\vec{\bar{x}}, \vec{x}) d\vec{\bar{x}} da,
	\label{eq:appendix_starting_point}
\end{equation}
which describes the transformation of the initial state $\mathcal{T}(\vec{x})$ under the action of the partial operations $K_{a}(\vec{\bar{x}}, \vec{x})$, defined in \eqref{eq:Ki_definition} (in the case of discrete indices, the integration over coordinates should be replaced by summation). First note, that integrating the final tomogram $\mathcal{T}'(\vec{x})$ over $X$ should give one as long as it is a probability distribution, hence
\begin{equation}
	\int \mathcal{T}(\vec{\bar{x}}) \Big( \int K_{a}(\vec{\bar{x}}, \vec{x}) dX da \Big) d\vec{\bar{x}} = 1,
\end{equation}
where we rearranged the integration order. For this equation to be fulfilled with an arbitrary initial state, the term enclosed in parentheses must be equal to $e^{i\bar{X}} \delta(\bar{\mu}) \delta(\bar{\nu})$\footnote[1]{Indeed, the latter expression is nothing but a dual symbol of the unit operator, namely $f_{\hat{1}}^{(d)}(\vec{x}) \equiv \operatorname{Tr} \{ \hat{1} \hat{\mathcal{D}}(\vec{x}) \}$. So, it is easy to see that convolving this symbol with an arbitrary tomogram leads to $\int \mathcal{T}(\vec{x}) f_{\hat{1}}^{(d)}(\vec{x}) d\vec{x} = (\mathcal{T}, f_{\hat{1}}) = \operatorname{Tr} \{ \rho \hat{1} \} = 1$.}, so the kernels of partial operations in probability representation must satisfy
\begin{equation}
	\int K_{a}(\vec{\bar{x}}, \vec{x}) dX da = e^{i\bar{X}} \delta(\bar{\mu}) \delta(\bar{\nu}).
\end{equation}
Substituting \eqref{eq:Ki_definition} into this formula, we eventually get the desired condition \eqref{eq:completeness_condition}. In particular, integrating \eqref{eq:completeness_condition} over $\bar{\mu}$ and $\bar{\nu}$, this relationship reduces to a weaker constraint
\begin{equation}
	\frac{1}{(2\pi)^2} \int f_{a}(X_1, \mu_1, \nu_1) f_{a}^* (X_2, \mu_1, \nu_1) e^{i(X_1 - X_2)} dX_1 dX_2 d\mu_1 d\nu_1 da = 1,
\end{equation}
which, being written using scalar product \eqref{eq:symbols_product}, is just $\int (f_{a}, f_{a}) da = 4\pi^2$. Note, that the same result could be obtained were we chosen a straightforward way that involves the recovering of the density matrix from \eqref{eq:appendix_starting_point} and its subsequent normalization, $\operatorname{Tr} \hat{\rho}' = 1$.

%\section*{References}

\bibliographystyle{ieeetr}
\bibliography{refs}

\end{document}